\documentclass{mn2e}
\usepackage{psfig}

\def\ltsima{$\; \buildrel < \over \sim \;$}
\def\lsim{\lower.5ex\hbox{\ltsima}}
\def\gtsima{$\; \buildrel > \over \sim \;$}
\def\gsim{\lower.5ex\hbox{\gtsima}}
\def\be{\begin{equation}}
\def\ee{\end{equation}}

\begin{document}

\title[small-scale magnetic field theory] {Gamma-Ray Burst Afterglow emission with a decaying magnetic field}

\author[Rossi \& Rees]
{Elena Rossi \& Martin J. Rees \\
Institute of Astronomy, University of Cambridge, Madingley Road,
Cambridge CB3 0HA, England \\ 
{\tt e-mail:emr,mjr@ast.cam.ac.uk}}

\maketitle
\begin{abstract}
In models for gamma ray burst afterglows, it is normally assumed that
the external shock strongly amplifies the magnetic field and that this
field maintains a steady value throughout the shocked region.  We
discuss the effects of modifying this (probably simplistic) assumption
by allowing for a short decay time.  The observations are incompatible
with a post-shock field that decays too rapidly.  However if the field
pervades only a few percent of the total thickness of the shocked
shell (and the electrons undergo only inverse Compton losses in the
remainder) the model could be compatible with data.  This would
suggest a strong dependence of model parameters on the uncertainties
of shock physics, therefore calling for independent external density
estimates.  We claim that afterglow emission should be instead seen as
a laboratory where we can understand relativistic shock physics.  The
model we propose here together with afterglow observations in all
wavebands could help to pin down the field structure.

\end{abstract}


\section{Introduction}

It is widely accepted that GRBs afterglow emission is due to
non-thermal radiation mechanisms produced in relativistic shocks, as
the ejected matter from a highly energetic explosion expands into the
surrounding medium.  With current facilities we can observe at photon
energies up to 10 keV and the modeling of the observed lightcurves
shows in general a predominance of synchrotron emission over inverse
Compton scattering on the synchrotron photons.  (Panaitescu \& Kumar
2001 (thereafter PK01)).  Only in one case, GRB000926, has a Compton
component been inferred (Harrison et al. 2001). This fact sets an upper
limit on the density of the medium surrounding the explosion, whose
properties can help us to unveil GRBs progenitors.  In the standard
framework the dynamics of the remnant expansion follows an adiabatic
law.  This behavior is expected when the electrons cannot cool on the
expansion timescale or (even when they can) the bulk of the energy
remains in relativistic protons and magnetic fields.  The post shock
magnetic field, necessary for the synchrotron emission, is mainly
generated during the shock itself, because the shock compression of
the pre-existing field alone would lead to a negligible magnetic
energy per particle (e.g. Gruzinov 2001).  The magnetic field is
normally assumed to be uniform and its lengthscale to be of the order
of the remnant scale. The electron energy distribution behind the
expanding shock front is given by a power law, down to a minimum
injection energy.  Modeling eight afterglow lightcurves within this
framework has resulted in a wide range of density from $\sim10^{-3}$
to $\sim27$ cm$^{-3}$, strongly indicating a low density environment
(PK01).  This is inconsistent with other independent observations
placing GRBs birth in a star forming region: SN bumps in afterglow
lightcurves (e.g. Bloom et al. 1999, Lazzati et al 2001, Bloom et
al. 2002), metal lines (Antonelli et al 2000, Piro et al 2000, Reeves
et al 2002), afterglow localizations close to star forming regions
(Bloom, Kulkarni, Djorgovski 2002, Jaunsen et al 2002) and estimations
of high density environment from the temporal evolution of external
column density (Lazzati \& Perna 2002). Another remarkable inference
from this model is that the fraction of internal energy given to
electrons ($\epsilon_e$) is always much larger then the fraction given
to magnetic field ($\epsilon_b$). This implies that the magnetic field
is relatively weak and the Compton emission dominates the overall
cooling of the electrons.

Recently many publications have investigated the somewhat poorly
understood generation of magnetic field in strong relativistic
collisionless shocks (e.g. Kazimura et al 1998, Gruzinov and Waxman
1998, Medvedev \& Loeb 1999, Gruzinov 2001).  Kazimura et al (1998)
found generation of a small-scale quasi-static magnetic field; similar
results were found by Gruzinov (2001) at small simulation times but
his longer run shows that the field quickly decays by Landau damping;
as a result synchrotron emission could be insignificant.  Medvedev \&
Loeb (1999) found more reassuring results.  Relativistic two-stream
magnetic instability can, they claim, generate stable randomly
oriented strong magnetic field in the plane of the collisionless shock
front, so synchrotron emission is possible.  These contradictory
results show our ignorance of the possible acceleration process, and
of how (or if) a stable magnetic field is created there.  In the
standard model this ignorance is ''parameterized `` through
$\epsilon_e$ and $\epsilon_B$: the simplistic assumption is made that
a persistent magnetic field makes these parameters uniform in space
and time.  An observation that maybe suggests a time dependence of
$\epsilon_e$ and $\epsilon_B$ is the independent determination of
these parameters in GRB 970508 afterglow by Wijers and Galama (1999)
at 12 days and Frail, Waxman and Kulkarni (2000) at $\sim 1$ year:
while the former found $\epsilon_e=0.12$ and $\epsilon_B=0.089$ the
latter inferred $\epsilon_e\simeq \epsilon_B=0.5$.

These uncertainties, coupled with the apparent problems with the
standard model, motivate us to explore the consequences for afterglow
emission of a different scenario, where the propagation of the
magnetic field to large scale is disfavored.  In this contest we
particularly focus on the possibility that external density could be
larger, keeping the total energy to a reasonable value, without
loosing a general agreement with data.  In fact, at this stage, we
deal with sparse set of measurements, almost never simultaneous at all
wavelengths. These data can be modeled in different frameworks,
(e.g. GRB 970508 (Frail et al 2000, Chevalier \& Li 2000), GRB 980519
(Frail et al 2000b, Chevalier \& Li 1999), GRB 000301C (Berger et al
2000, Li \& Chevalier 2001), GRB 991208 (Galama et al 2000, Li \&
Chevalier 2001 and GRB 000418 (Berger et al 2001)).  This analysis is
meant to investigate how robust are the current estimation of shock
and density parameters.

In this paper we present this more general theory for the GRB
afterglow emission (\S 2); our conclusion (\S 3) is that a qualitative
comparison of this model with data suggests the possibility that a
shorter magnetic lengthscale can be involved, affecting the estimates
of parameters, but only a quantitative modeling of data can give the
conclusive answer.

\section{Short magnetic lengthscale theory}

The assumption of a post shock magnetic field persisting during all
the expansion time has three main consequences for the synchrotron (S)
emission and spectrum.  First the emitting region linear dimension is
$R\simeq c \times t_{exp}$, where $t_{exp}$ is the expansion time in
the lab frame; this enters in the calculation of the observed peak
flux $F_p$.  Then the observed cooling frequency $\nu_c$ is computed
using the Lorenz factor, $\gamma_c$, of the electrons that cool
radiatively on a timescale equal to the remnant age. Finally the
lengthscale for synchrotron self-absorption has to be compared with
the dimension of the fireball in order to compute the observed
absorption frequency $\nu_a$.  The only break frequency which is not
affected by any assumption about the extension of the magnetic field,
is the observed peak frequency $\nu_m$, corresponding to the minimum
random Lorentz factor $\gamma_m$: it depends crucially only on the
fractions of internal energy given to electrons, $\epsilon_e$, and
given to the magnetic field $\epsilon_b$. Throughout the entire
emission volume electrons cool also via inverse Compton (IC) on the
synchrotron photons (Self Synchrotron Compton: SSC).  This cooling
process dominates the electrons cooling if the Compton parameter
$Y=L_{IC}/L_S$ (IC and S luminosity ratio) is greater than one. This
happens if $\epsilon_b \ll \epsilon_e$ and current estimates find in
most objects $\epsilon_b/\epsilon_e\sim 10^{-1}-10^{-2}$, (Wijers \&
Galama 1999, Granot et al 1999, PK01).
\begin{figure}
\parbox{0.49\textwidth}{\psfig{file=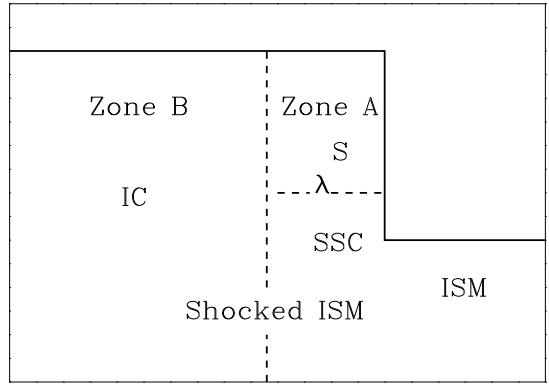,width=0.48\textwidth}}
\caption{{A schematic picture of the emitting region (shocked ISM). 
In region A, just behind the shock front, there are magnetic field and
newly accelerated electrons that emit synchrotron and SSC for a time
$\lambda/c$ before downstreaming into the (unmagnetized) region B
where they Compton scatter the S radiation coming from zone A.}}
\label{fig:car}
\end{figure}

If the magnetic field behind the shock persists for an average
timescale of $t_b<t_{exp}$, its lengthscale is $\lambda\simeq c \times
t_b<R$.  In principle $\lambda$ could depend on time, however we do
not have any consolidated understanding of this process, therefore for
simplicity we will assume throughout this paper that
$\lambda/R=\delta$ is constant during the afterglow phase.  At each
time the emitting region is then divided in two zones (see
Fig~\ref{fig:car}).  The first zone (region A) is the layer with
dimension $\lambda $ just behind the shock front, where there are
newly accelerated electrons and magnetic field: here electrons emit
via S and SSC; the second zone (region B) of linear dimension
$R-\lambda$, is filled by electrons that have just cooled by S and SSC
emissions and that now Compton scatter the radiation coming from the
first layer.  We suppose that the newly accelerated electrons are
advected with the flow; they produce synchrotron radiation only for a
short time (with a cooling break therefore only at very high energies)
in region A near the shock front and then they emit most of the IC
radiation while downstreaming through region B. On the other hand if
the electrons diffused freely, so that they could move back into the
magnetized region after having lost most of their energy via IC, then
the synchrotron spectrum would have a lower-energy break. This is
probably less likely therefore we concentrate on the first case. The
total observed spectrum then results from the sum of region A and
region B spectra.

\subsection{region A: S and SSC emission}
The consequences of $\lambda<R$ on the S spectrum in region A  
can be derived easily from what we mentioned before.
The observed peak frequency $\tilde{\nu_m}$ 
remains unchanged,
\be
\tilde{\nu_m}=\nu_m,
\label{eq:num}
\ee

and 

\be
\tilde{\gamma_m}\propto\tilde{\nu_m}^{1/2} =\gamma_m.
\label{eq:gm}
\ee
In our generalized model with $\delta<1$ we denote quantities by the
symbol '$\tilde{q}$', where '$q$' is defined for $\delta=1$, (for the
explicit definitions of quantities in the standard model we refer to
Panaitescu \& Kumar 2000).

The comoving peak intensity is $I'\propto \tilde{n}'B'\lambda'$, where
$\tilde{n}'$, $B'$ and $\lambda'$ are respectively the external
density, the magnetic field and the emitting region linear dimension
in the comoving frame; therefore the observed peak flux $\tilde{F_p}$
is smaller than $F_p$ by a factor $\delta$:

\be
\tilde{F_p}\simeq\Gamma^{3} (\tilde{n}'B'\lambda') \frac{R^{2}}{\Gamma^{2}}=\delta\times F_p,
\label{eq:fp}
\ee

\noindent
where $\Gamma$ is the bulk Lorentz factor.  Since the magnetic field
persists for a shorter time, a smaller fraction of electrons cool
significantly, before the magnetic field disappears in that region.
The Lorentz factor of the electrons, whose cooling time is equal to
the typical timescale of the system, $\tilde{\gamma_c}$ is therefore
higher:
\be
\tilde{\gamma_c}\propto \frac{1}{t_b'\,B'^{2}\,(1+\tilde{Y})}=\gamma_c \times {\cal Y}\,\delta^{-1}
\label{eq:gc},
\ee

\noindent
where ${\cal Y}=(1+Y)/(1+\tilde{Y})$.  The observed cooling frequency
$\tilde{\nu_c}\propto B'\,\tilde{\gamma_c}^{2}\,\Gamma$ is then

\be
\tilde{\nu_c}=\nu_c\times {\cal Y}^2\,\delta^{-2}.
\label{eq:nuc}
\ee

The synchrotron self-absorption optical thickness can be approximated
as

\be
\tilde{\tau_a}(\nu)\propto \left(\frac{\tilde{n}\,\lambda}{B'\tilde{\gamma_p}^{5}}\right)\left(\nu/\tilde{\nu_p}\right)^{-5/3}=\tilde{\tau_p}\left(\nu/\tilde{\nu_p}\right)^{-5/3},
\label{eq:tau}
\ee

\noindent 
for $\nu<\tilde{\nu_p}$, where $p=c$ when the electrons are radiative
($\tilde{\gamma_c}<\tilde{\gamma_m}$) and $p=m$ when they are
adiabatic ($\tilde{\gamma_c}>\tilde{\gamma_m}$).  We will refer to the
first case as fast-cooling regime and to the second case as
slow-cooling regime.  The absorption frequency corresponds to
$\tilde{\tau_a}(\tilde{\nu_a})=1$ and in slow cooling regime is given
by

\be
\tilde{\nu_a}=\nu_a \times \delta^{3/5},
\label{eq:nuas}
\ee

\noindent
where we use Eq.~\ref{eq:num} and Eq.~\ref{eq:tau}; in fast cooling
regime Eqs ~\ref{eq:nuc}, ~\ref{eq:tau} and ~\ref{eq:yf} (${\cal
Y}=1$) give

\be
\tilde{\nu_a}=\nu_a \times \delta^{8/5}.
\label{eq:nuaf}
\ee

\noindent
Eqs~\ref{eq:fp},~\ref{eq:nuas} and ~\ref{eq:nuaf} are calculated in
the case $\tilde{\nu}_a<\tilde{\nu}_p$ (see Granot \& Sari 2001 for
the different possible S spectra).  Nevertheless $\tilde{\nu}_a
\ge\tilde{\nu}_p$ for high densities
\be
 \tilde{n}\ge 7\times 10^{2} \delta^{-12/7} E_{53}^{-12/10}
(1+\tilde{Y})^{-3} \epsilon_{b,-2}^{-27/10}
\label{nlimif}
\ee

\noindent
in fast cooling and in slow cooling
\be
\tilde{n}\ge 8.2\times 10^{5}\delta^{-1} E_{53}^{-1/2}\epsilon_{b,-2}^{1/2}\,\epsilon_{e,-2}^{5}
\, (1+z)\,t_{day}^{5/2}
\label{nlimis}
\ee

\noindent
where $E$ is the isotropic equivalent energy and $t_{day}$ is the
observed time in days.  The inferred external density is reasonable
for $\delta=0.1$ ($\tilde{n}\ge 3.6 \times10^{4}$cm$^{-3}$ and
$\tilde{n}\ge 8.5\times10^{6}\,t_{day}^{5/2}$cm$^{-3}$ respectively)
and for $\delta=0.01$ ($\tilde{n}\ge 2\times 10^{6}$cm$^{-3}$ and
$\tilde{n}\ge 8.5\times 10^{7}\,t_{day}^{5/2}$cm$^{-3}$) but for lower
$\delta$ the density lower limits expressed in Eq~\ref{nlimif} and
Eq~\ref{nlimis} are perhaps too high for the afterglow environment.
For simplicity we focus thereafter on $\tilde{\nu}_a<\tilde{\nu}_c$
but the reader should keep in mind that these cases hold as long as
Eq~\ref{nlimif} and Eq~\ref{nlimis} are not satisfied.

The magnetic energy $\frac{B'^{2}}{8\pi}\propto n \Gamma^{2}$ and the
Compton parameter are higher at the beginning of the afterglow
evolution and so the electrons are more likely then to be radiative.
Since the minimum injected $\gamma_m\propto \Gamma$ decreases with
time while $\gamma_c$ increases as the efficiency of radiative loss
decreases, the electrons undergo a transition to the adiabatic regime.
The transition time between the fast cooling and the slow cooling
regime is
\be
\tilde{T}_{fs}=Tfs\times \delta^2;
\ee

\noindent
for 
\be
\tilde{n}<40\,\delta^{-2}\,E_{53}^{-1}\,(Y_F+1)^{-2}\,\epsilon_{B,-2}^{-2}\epsilon_{e,-1}^{-2}\,(1+z)^{-1}\,t_{day}^{-1}
\ee
the slow-cooling regime dominates the fireball evolution during times
when the afterglow is observed ($\gsim 1$ day).

The accelerated electrons that upscatter synchrotron photons give rise
to a SSC component in the spectrum (for the SSC in the standard model
see Sari \& Esin 2000).  During the afterglow phase the medium has a
Thomson optical depth
\be
\tilde{\tau}=\frac{1}{3}\,\tilde{n}\,\sigma_T\,R\,\delta
\label{tau}
\ee
\noindent
of the order of $\tilde{\tau}\simeq 10^{-5}\times \delta$ for
$\tilde{n}=10^{2}$cm$^{-3}$ ($\sigma_T$ is the Thomson cross section),
therefore only a minor fraction of the emission is upscattered, and
the synchrotron flux is not significantly altered.  Nevertheless if $
\tilde{Y}\ge 1$ the electron cooling is Compton dominated and SSC
upscatters low energy photons, that might be detected as a Compton
component in x-ray and, in the future, in higher energy band spectra
(Sari, Narayan \& Piran 1996).  The parameter $\tilde{Y}$ can be
defined as the mean number of scatterings times the average fractional
energy change per scattering, gained by a photon traversing a finite
medium (Rybicki \& Lihgtman 1979).  For a Thomson thin, relativistic
medium $\tilde{Y}$ is

\be
\tilde{Y}\simeq \frac {4}{3}\tilde{\tau}\,\, \int {\tilde{\gamma}^2
 N(\tilde{\gamma}) d\tilde{\gamma}}
\label{eq:y}
\ee

\noindent
where $ N(\tilde{\gamma})$ is the normalized $\tilde{\gamma}$
distribution (see e.g. PK00 for explicit formulae for $\nu_a<\nu_p$).
For $\tilde{Y} > 1$ one may need to consider more than one scattering
per photon; however the energy of a photon in the rest frame of the
second scattering electron is a factor $\gamma^{3}$ higher than the
original energy, generally exceeding the Thompson regime limit (511
keV); the scattering cross section is then substantially reduced
(Klein-Nishina (KN) regime) and further scatterings inhibited.  As a
consequence we will consider only single scattering of S photons.  The
KN effect limits also the electrons that mostly contribute to the SSC
luminosity. They have a Lorentz factor
\be
\tilde{\gamma} \le \tilde{\gamma}_{KN}\simeq 511\,\,\Gamma / h\,\tilde{\nu}_{Ep},
\label{eq:KN}
\ee

\noindent
where $\tilde{\nu}_{Ep}$ is the S energy peak frequency
($\tilde{\nu}_c$ in slow cooling and $\tilde{\nu}_m$ in fast cooling).
In zone A the S spectrum can have a very high $\tilde{\nu_c}$ (see
Eq~\ref{eq:nuc}); consequently in slow cooling $\tilde{\gamma}_c$
could be greater than $\tilde{\gamma}_{KN}$ and the Klein-Nishina
effect substantially decreases $\tilde{Y}$ (we will refer to this
situation as ``zone A in KN regime'').  Moreover if $\tilde{Y}\simeq
1$ (which is the case in most simulations) we can omit the
contribution from $\tilde{\gamma}>\tilde{\gamma}_{KN}$.
Eq.~\ref{eq:y} for $2<p<3$ gives then

\be
\tilde{Y_S}\,(1+\tilde{Y_S})^{(3-p)}=Y_S\,(1+Y_S)^{(3-p)}\times \delta^{(p-2)},
\label{eq:ys1}
\ee 

\noindent
for $\tilde{\gamma}_c\le\tilde{\gamma}_{KN}$ and

\be
\tilde{Y_S}\,(1+\tilde{Y_S})^{(2(p-3))}=
Y_S\,(1+Y_S)^{(2(p-3))}\times \delta^{(7-3p)},
\label{eq:ys2}
\ee 

\noindent
for $\tilde{\gamma}_c\le\tilde{\gamma}_{KN}$.
If the electrons are in the fast cooling regime it is very unlikely that 
$\tilde{\gamma}_m>\tilde{\gamma}_{KN}$ and Eq.~\ref{eq:y} gives

\be
\tilde{Y_F}=\frac {4}{3}\,\tilde{\tau}\,\tilde{\gamma}_{m}\,\tilde{\gamma_c}=Y_F;         
\label{eq:yf}
\ee
note that $\tilde{Y_F}$ does not depend on $\delta$, because most of
the electrons cool before the field disappears.  Given a S seed
spectrum $\tilde{F}_{S}(\nu_s)$, the resulting single scattering IC
spectrum is (Rybicki \& Lightman, 1979)
\be
\tilde{F_{\nu}}^{IC}=A\,\int_{0}^{1} \tilde{F}_{S}
 \left(\frac{\nu}{4\tilde{\gamma}^{2}X}\right)\,\,f(X)\, 
dX\,\int_{\tilde{\gamma}_1}^{\tilde{\gamma}_2} N(\tilde{\gamma})\, 
d\tilde{\gamma},
\label{eq:ficrl}
\ee

\noindent
where $A=3\sigma_TR\delta$ (in this model),
$X=\nu/4\tilde{\gamma}^{2}\nu_s$ and $f(X)=2X\ln{X}+X-2X+1$ includes
the exact cross section angular dependence in the limit
$\tilde{\gamma}\gg1$ (Blumenthal and Gold,1970). For the cases we are
treating it is possible to use for the first integral the analytic
expressions given by Sari \& Esin (2000) in their Appendix A.  The IC
spectral peak flux and break frequencies for $\delta<1$ are related to
the IC spectrum for $\delta=1$ in the following way
\be
\tilde{F_p}^{IC}\simeq \tilde{\tau} \tilde{F_p}= F_p^{IC} \times \delta^{2}
\label{eq:fpic}
\ee
using Eq~\ref{eq:fp} and Eq~\ref{eq:tau}
\be
\tilde{\nu}_m^{IC}\simeq 2\tilde{\gamma_m}^2\,\tilde{\nu_m}=\nu_m^{IC},
\label{eq:nuicm}
\ee
 using Eq.~\ref{eq:gm},~\ref{eq:num};
\be
\tilde{\nu}_c^{IC}\simeq 2\tilde{\gamma_c}^2\,\tilde{\nu_c}=\nu_c^{IC}\times \left({\cal Y}\,\delta^{-1}\right)^{4},
\label{eq:nuicc}
\ee
using Eq.~\ref{eq:gc},~\ref{eq:nuc};
\be
\tilde{\nu}_a^{IC}\simeq 2\tilde{\gamma_m}^2\,\tilde{\nu_a}=\nu_a^{IC} \times \delta^{3/5},
\label{eq:nuicas}
\ee
in the slow cooling regime using Eq.~\ref{eq:gm},~\ref{eq:nuas} and 
\be
\tilde{\nu}_a^{IC}\simeq 2\tilde{\gamma_c}^2\,\tilde{\nu_a}=\nu_a^{IC} \times \,\,\delta^{-2/5}
\label{eq:nuicaf}
\ee
in the fast cooling regime, using Eq.~\ref{eq:gc},~\ref{eq:nuaf}.

\subsection{Region B: IC emission}
Zone B extends over a distance $(R-\lambda)$; since we are mainly
interested in $\delta \le 10^{-1}$, we approximate $(R-\lambda)\simeq
R$ throughout this paper.  This region is populated by electrons that
have radiatively cooled via S and SSC for a time $t_b$.  If they were
originally injected at the front shock with a power-low energy
distribution $N(\tilde{\gamma})\propto \tilde{\gamma}^{-p}$ then the
steady state distribution in region B depends on whether the electrons
were adiabatic or radiative in region A.  The $\tilde{\gamma}_{cB}$
for electrons that cool in a dynamical timescale $t_{exp}$ is in fact

\be
\tilde{\gamma}_{cB}=
\frac{m_e\,c}{\frac{4}{3}\sigma_T\,\frac{U_{rad}}{2}\,t_{dyn}},
\label{eq:gcb1}
\ee

\noindent
where $U_{rad}$ is the S radiation energy density and the factor $1/2$
takes into account that only half of the S photons travel from region
A towards region B in the electrons rest frame.  The calculation of
$U_{rad}$ in the two different regimes leads to
\be
\tilde{\gamma}_{cB}\propto\left\{\begin{array}{ll}
(n^{2}\,e_b\Gamma\tilde{\gamma}^{p-1}\,\tilde{\gamma}_c^{3-p}\,R^{2}\delta)^{-1}& \tilde{\gamma}_m<\tilde{\gamma}_c\\
 (n^2\,e_b\Gamma\tilde{\gamma}_m\,\,\tilde{\gamma}_c\,R^{2} \delta)^{-1.}& \tilde{\gamma}_m>\tilde{\gamma}_c
	\end{array}\right.
\label{eq:gcb}
\ee

\noindent
If electrons in region A are in the slow cooling regime, the new
steady state configuration is

\be
N(\tilde{\gamma})\propto\left\{\begin{array}{ll}
 \tilde{\gamma}^{-p} & 
 \tilde{\gamma}_m<\tilde{\gamma}<\tilde{\gamma}_{cB}\\
 \tilde{\gamma}^{-(p+1)} & 
 \tilde{\gamma}_{cB}<\tilde{\gamma}<\tilde{\gamma}_{c}\\
 \tilde{\gamma}^{-(p+2)} & \tilde{\gamma}>\tilde{\gamma}_{c}\\
 \end{array}\right.
 \label{eq:NgBs}
\ee
for slow cooling 
($\tilde{\gamma}_m<\tilde{\gamma}_{cB}$) or
\be
N(\tilde{\gamma})\propto\left\{\begin{array}{ll}
 \tilde{\gamma}^{-2} & 
 \tilde{\gamma}_{cB}<\tilde{\gamma}<\tilde{\gamma}_{m}\\
 \tilde{\gamma}^{-(p+1)} & 
 \tilde{\gamma}_{m}<\tilde{\gamma}<\tilde{\gamma}_{c}\\
 \tilde{\gamma}^{-(p+2)} & \tilde{\gamma}>\tilde{\gamma}_{c}\\
 \end{array}\right.
 \label{eq:NgBf}
\ee
for fast cooling ($\tilde{\gamma}_m>\tilde{\gamma}_{cB}$).
$\tilde{\gamma}_{cB}> \tilde{\gamma}_{c}$ is very unlikely, even if
the electrons in region A have two sources of energy loss (S and SSC),
because $t_b<t_{exp}$ and SSC can be inhibited further for KN effect.
If electrons in region A undergo fast cooling, the most common case is
$\tilde{\gamma}_{cB}< \tilde{\gamma}_{c}$ and the new steady state
configuration is
\be
N(\tilde{\gamma})\propto\left\{\begin{array}{ll}
 \tilde{\gamma}^{-2} & 
 \tilde{\gamma}_{cB}<\tilde{\gamma}<\tilde{\gamma}_{c}\\
 \tilde{\gamma}^{-3} & 
 \tilde{\gamma}_{c}<\tilde{\gamma}<\tilde{\gamma}_{m}\\
 \tilde{\gamma}^{-(p+2)} & \tilde{\gamma}>\tilde{\gamma}_{m}\\
 \end{array}\right.
 \label{eq:NgBff}
\ee

The ``external Compton'' spectrum in zone B is calculated by means of
Eq~\ref{eq:ficrl} with $A=\frac{1}{2} \sigma_T\,R$ and the Compton
parameter $\tilde{Y}_B$ using Eq~\ref{eq:y} and
Eq~\ref{eq:NgBs}-~\ref{eq:NgBff} with $\tilde{\tau}_B=\frac{\tau}{2}$.
Even if zone B has a higher number of IC emitting electrons , the
Compton parameter can be lower than in region A: a higher fraction of
the electrons in region B cool efficiently and they are in average
less energetic then in region A, therefore increasing the density
parameter $\tilde{Y}_B$ decreases while $\tilde{Y}$ increases because
in zone A $\gamma_c$ moves towards low values and the KN effect is
less and less important (until the electrons become radiative and
$\tilde{Y}_F$ does not depend on density). In this situation the SSC
radiation can actually account for most of the energy that goes into
IC luminosity, (this is the case in Figs~\ref{fig:figs} and
\ref{fig:figf}
right panels); anyway the SSC usually dominates
the spectrum at high energy because it has a higher energy break. 
The peak flux is 

\be
\tilde{F_p}^{IC}(\nu)\simeq \tilde{\tau}_B \tilde{F_p},
\label{eq:fpicB}
\ee

\noindent
which is a factor $\delta^{-1}/2$ higher than $\tilde{F_p}^{IC}$ in
region A; while the spectral breaks are related to the S seed spectrum
in slow cooling as follows
\be
\tilde{\nu}_a^{IC}\simeq 2\tilde{\gamma}_m^2\,\tilde{\nu}_a,
\label{eq:nuicaBs}
\ee
\be
\tilde{\nu}_m^{IC}\simeq 2\tilde{\gamma}_m^2\,\tilde{\nu}_m,
\label{eq:nuicmBs}
\ee

\be
\tilde{\nu}_c^{IC}\simeq 2\tilde{\gamma}_{cB}^2\,\tilde{\nu}_c,
\label{eq:nuiccBs}
\ee
if region B is in the slow cooling regime as well, and

\be
\tilde{\nu_a}^{IC}\simeq 2\tilde{\gamma}_{cB}^2\,\tilde{\nu}_a,
\label{eq:nuicaBf}
\ee
\be
\tilde{\nu}_c^{IC}\simeq 2\tilde{\gamma}_{m}^2\,\tilde{\nu}_c,
\label{eq:nuiccBs}
\ee
\be
\tilde{\nu}_m^{IC}\simeq 2\tilde{\gamma}_{cB}^2\,\tilde{\nu}_m,
\label{eq:nuicmBs}
\ee
if B is in the fast cooling regime.
If S seed spectrum is in the fast cooling regime 
and $\tilde{\gamma}_{cB}< \tilde{\gamma}_{c}<\tilde{\gamma}_m$
\be
\tilde{\nu_a}^{IC}\simeq 2\tilde{\gamma}_{cB}^2\,\tilde{\nu}_a,
\label{eq:nuicaBf}
\ee
\be
\tilde{\nu}_c^{IC}\simeq 2\tilde{\gamma}_{cB}^2\,\tilde{\nu}_c,
\label{eq:nuiccBs}
\ee
\be
\tilde{\nu}_m^{IC}\simeq 2\tilde{\gamma}_{c}^2\,\tilde{\nu}_m.
\label{eq:nuicmBs}
\ee

\section{Discussion and conclusions}
\begin{figure*}
\parbox{0.49\textwidth}{\psfig{file=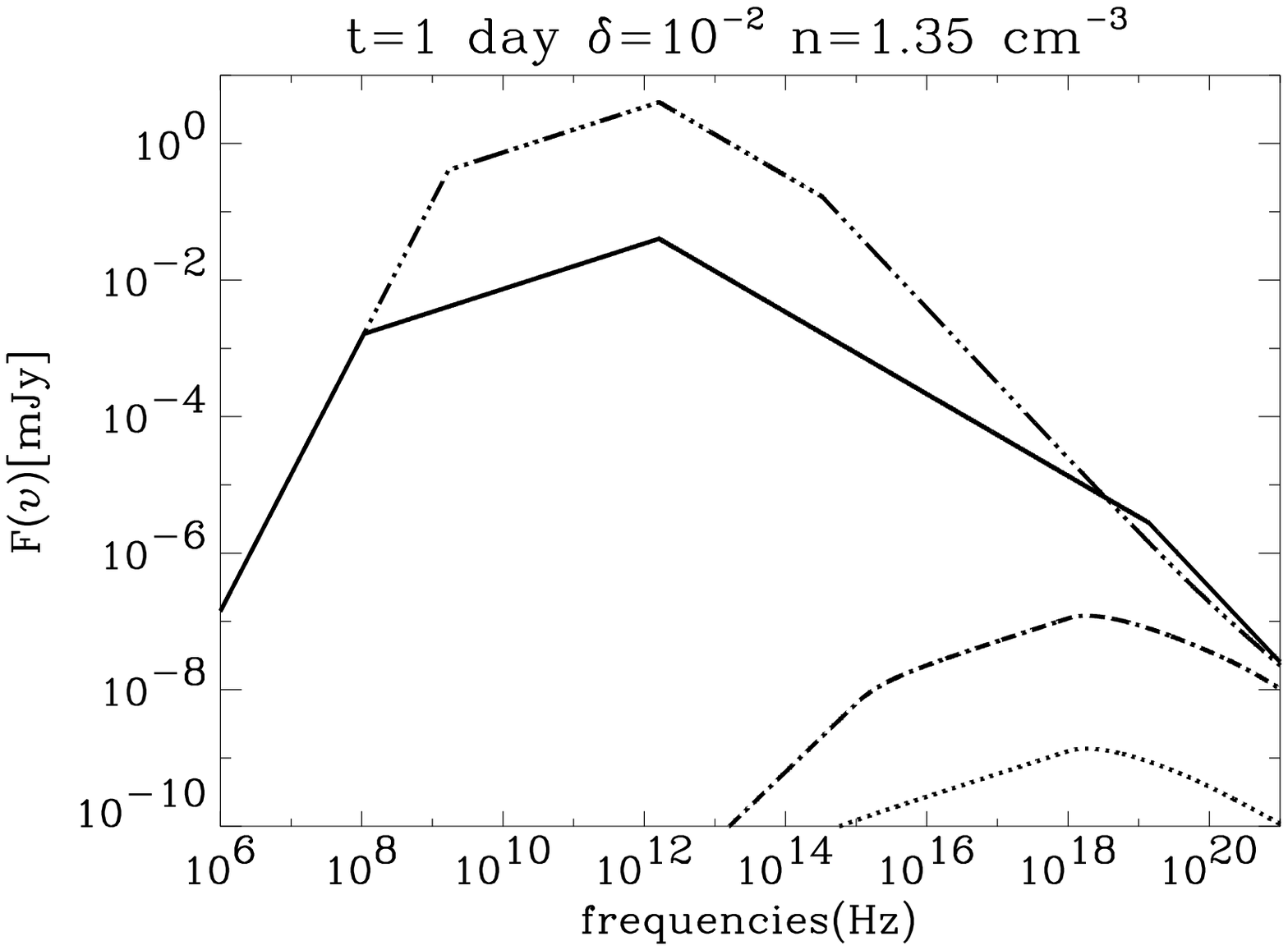,width=0.48\textwidth}}
\parbox{0.49\textwidth}{\psfig{file=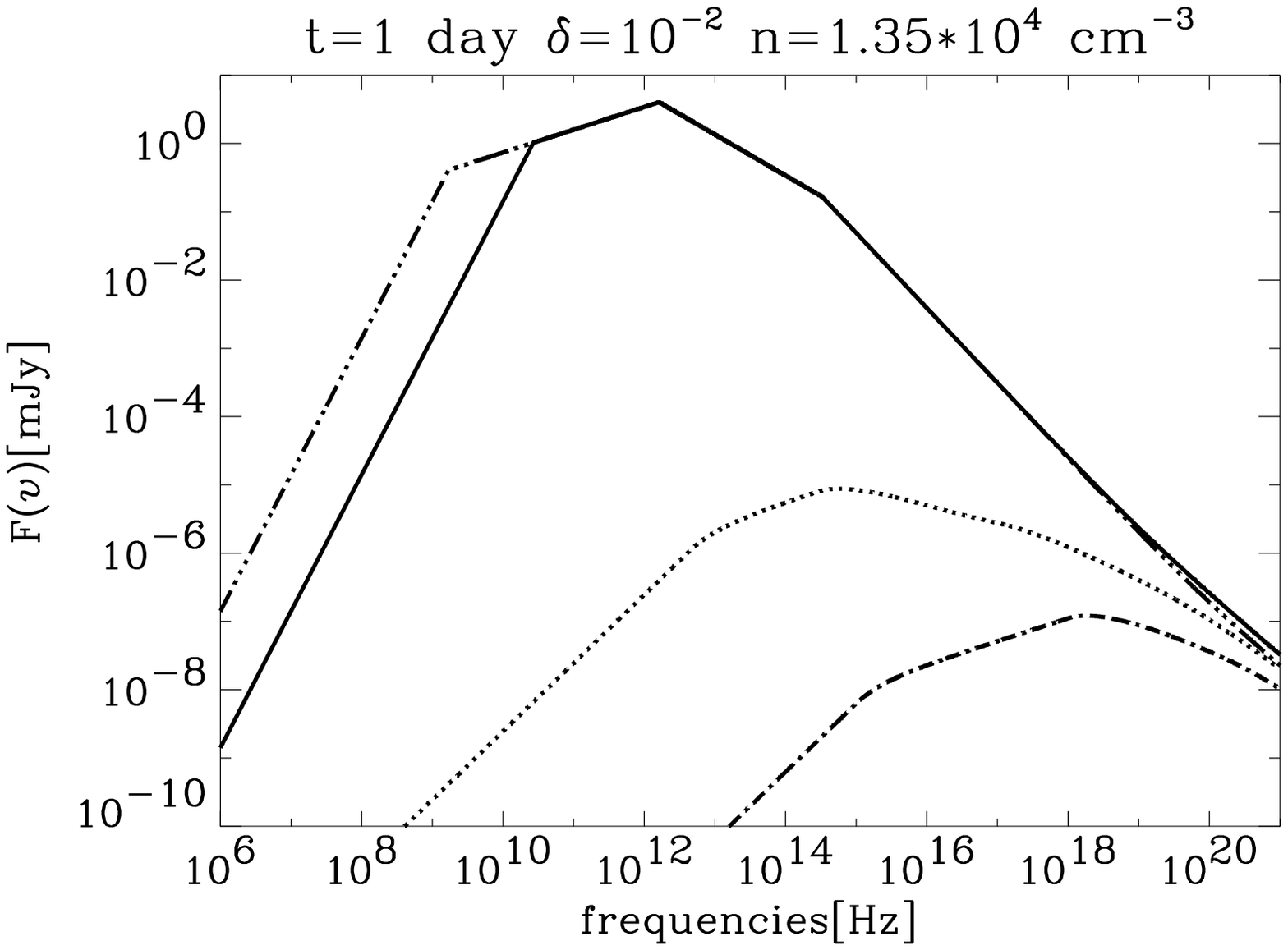,width=0.48\textwidth}}
\caption{{  An example of how the spectrum modifies if $\delta<1$.
Total (solid line) and SSC+IC (dot line) spectra for $\delta=0.01$ and
total (3 dots-dash line) and SSC (dot-dash line) spectra for
$\delta=1$ at 1 day.  In the left panel the two total spectra have all
parameters, except for $\delta$, the same: $n$=1.35 cm$^{-3}$, E=$3
\times10^{52}$ erg, $\epsilon_e=6 \times 10^{-2}$, $\epsilon_B=4
\times 10^{-3}$, $p$=2.2 at a redshift z=1.  These parameters are
typical of broadband afterglow lightcurves modeled in the standard
framework (PK01).  Both spectra are in the slow cooling regime (also
zone B with $\gamma_{cB}<\gamma_c$).  The SSC in zone A is highly
inhibited by the KN effect; therefore most of the IC component is due
to ``external Compton'' in zone B.  The resulting Compton parameters
are $\tilde{Y}=0.40$ and $Y=1.05$.  In the right panel we raise the
density in the $\delta=0.01$ spectrum until $\tilde{F_p}=F_p$,
$\tilde{n}=1.35\times10^{4}$ cm$^{-3}$.  Zone B is now in the fast
cooling regime therefore $\gamma_{cB}$ moves towards low values but
since in zone A the electrons are almost all in Thompson regime the
total $\tilde{Y}$ increases, $\tilde{Y}\simeq 1.10$: region A now
dominates the IC emission.  It is worth noticing that the two total
spectra are distinguishable only in the radio waveband where the
$\delta=0.01$ flux is more absorbed.}
\label{fig:figs}}
\end{figure*}

\begin{figure*}
\parbox{0.49\textwidth}{\psfig{file=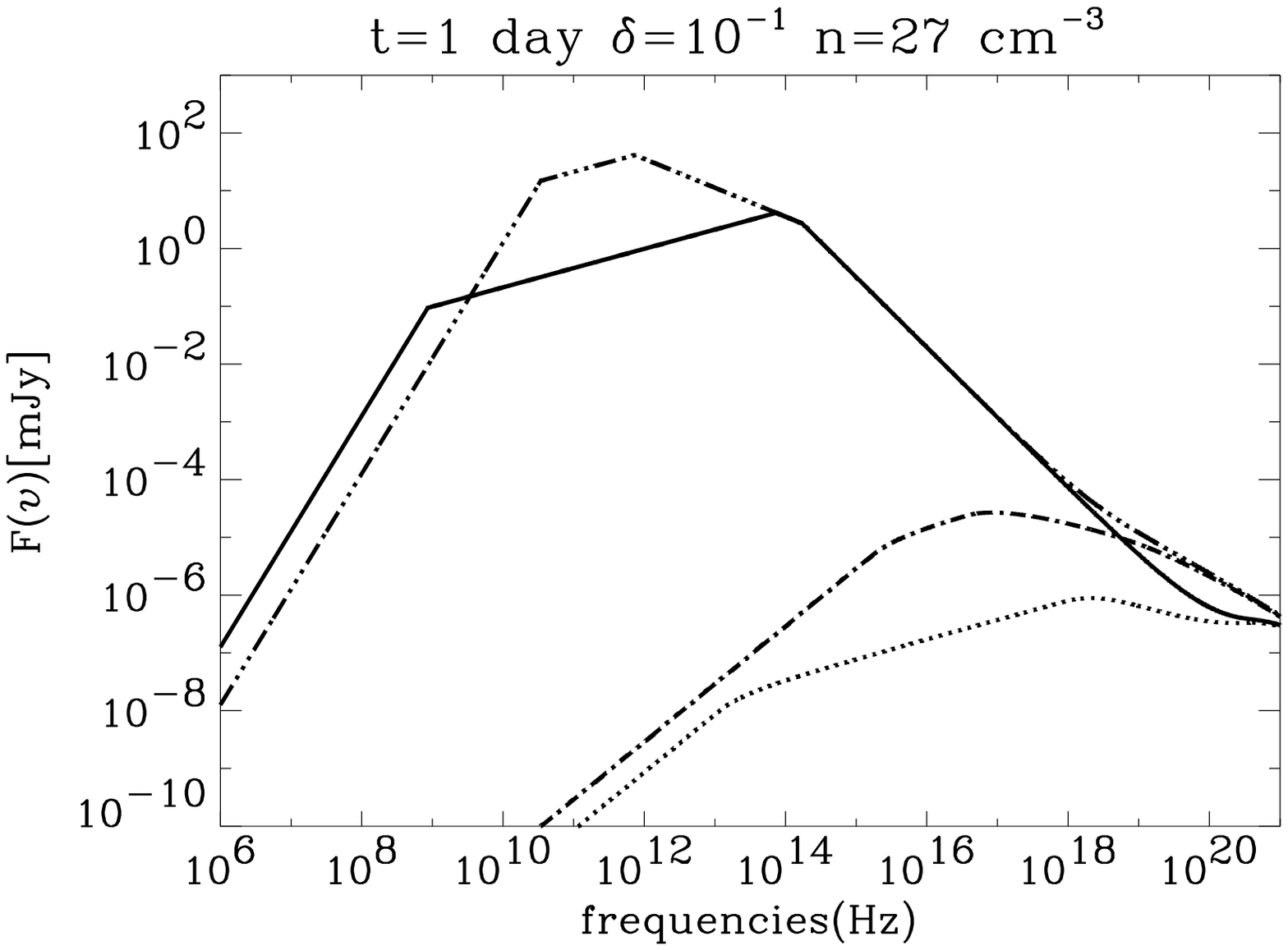,width=0.48\textwidth}}
\parbox{0.49\textwidth}{\psfig{file=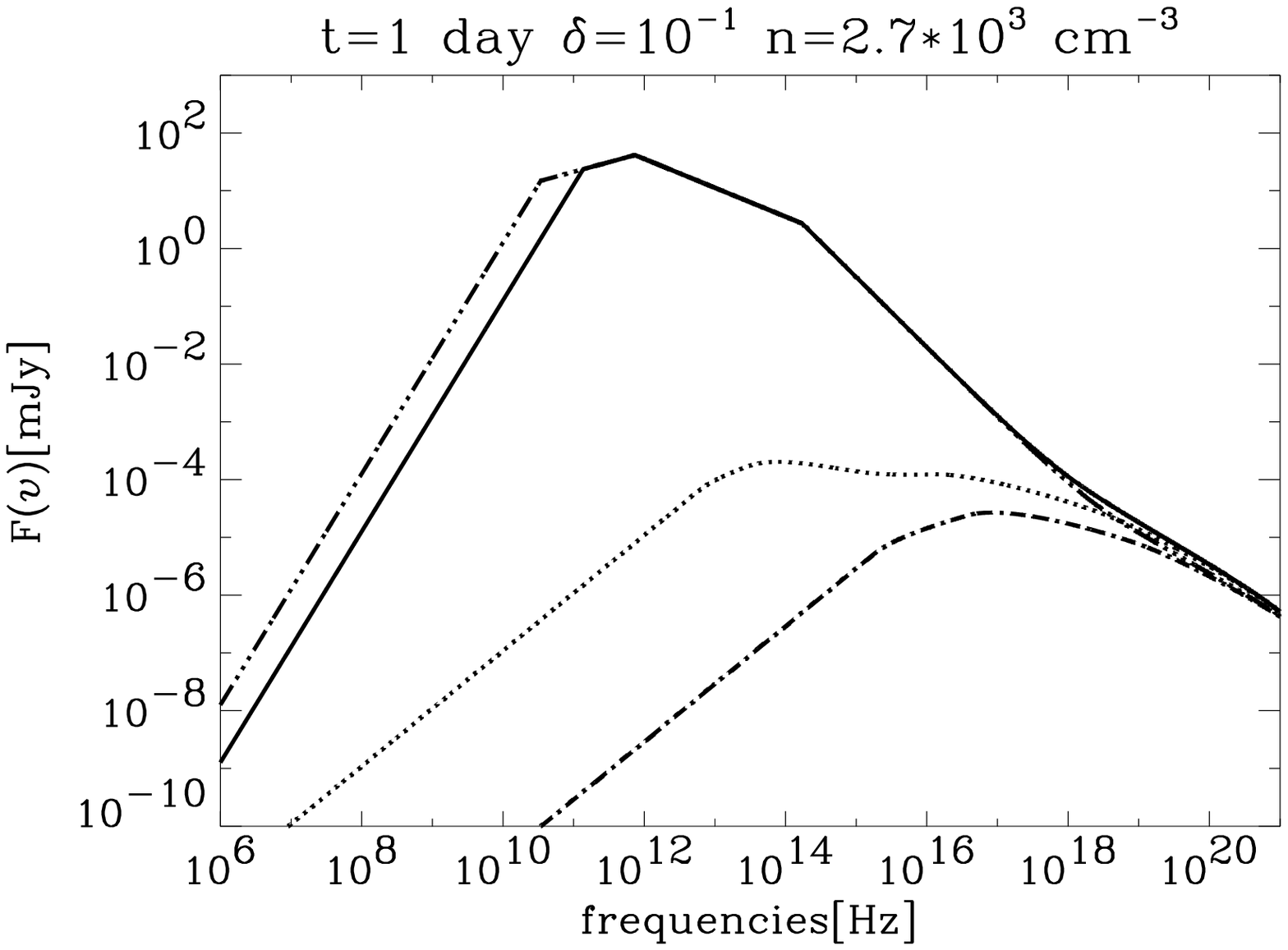,width=0.48\textwidth}}
\caption{{ Same as Fig.~\ref{fig:figs} with parameters  
$n=27$ cm$^{-3}$, $E=1.8 \times 10^{53}$ erg,$\epsilon_e=0.3$ and
$\epsilon_B=8\times10^{-3}$ and p=2.43 and $\delta=0.1$. This set of
parameters results from the best fit of 000926 (Harrison et al. 2001).
The redshift of this burst is $z=2.0639$.  In the left panel both
spectra are in fast cooling (also zone B with $\gamma_{cB}<\gamma_c$).
The Compton parameters are $\tilde{Y}=2.59$ and $Y_F=2.34$.  In the
right panel we rise the value of density in the $\delta=0.1$ spectrum
until $\tilde{F_p}=F_p$, $\tilde{n}=2.7\times10^{3} cm^{-3}$. The
spectrum is obviously still in fast cooling with a lower
$\gamma_{cB}$; therefore the total $\tilde{Y}$ is lower ,
$\tilde{Y}\simeq 2.37$ and the main contribution to the IC luminosity
comes from region A.  Again the two total spectra are distinguishable
only in the radio waveband.}}
\label{fig:figf}
\end{figure*}

In Fig~\ref{fig:figs} and Fig.~\ref{fig:figf} left panels, we show as
examples how the spectrum (respectively for fast and slow cooling)
modifies if a short magnetic lengthscale is taken into account.  Note
that S and IC peak fluxes are lower (see also Eq.~\ref{eq:fp},
Eq.~\ref{eq:fpic}, Eq.~\ref{eq:fpicB}), a lower energy break
corresponding to the absorption frequency is present (see
Eqs.~\ref{eq:nuas} and ~\ref{eq:nuaf}) and the cooling frequency is at
higher energies (see Eq.~\ref{eq:nuc}).  The SSC component in the slow
cooling spectra of region A (Fig~\ref{fig:figs}) is depleted because
the KN effect is important. In the right panels we raise the density
until $\tilde{F}_p\propto \tilde{n}^{1/2} \delta =F_p \propto
n^{1/2}$, thus $\tilde{n}=\frac{n}{\delta^{2}}$.  We obtain
$\tilde{\nu_c}=\nu_c$ and the $\delta<1$ spectrum fails to match the
standard spectrum only at radio wavelengths. In this band such high
densities, in fact, overcompensate the shorter magnetic lengthscale
and $\tilde{\nu_a}$ becomes greater then $\nu_a$.  Another important
feature is that the total (region A plus region B) $\tilde{Y}$
parameter is never much higher then $Y$, even when $Y$ corresponds to
a density $\delta^2$ lower; in fact when $\tilde{F}_p=F_p$, the main
contribution to IC luminosity comes from region A and Eq~\ref{eq:yf}
and Eq~\ref{eq:ys1} hold: only $\tilde{Y}_S$ depends on density and
$\delta$ but the dependence are very weak
($\tilde{Y}_S\,(1+\tilde{Y}_S)^{(3-p)}\propto
\tilde{n}^{(p-2)/2}\times \delta^{(p-2)}$).  These illustrative
examples of how the model works should now be followed by a modeling
of real data. In fact the general agreement between the two spectral
shapes suggests that data could also be well reproduced in this
framework, with possibly different values for the break frequencies
and peak flux respect to the standard model.  As regard the
corresponding estimated density compared to the standard model, there
are two competitive facts: the peak flux is brought to observed ranges
by an higher density but the absorption frequency cannot be too high
(see fig. 2 and 3) (note that unlike the standard model a high
density, $\gsim 10^{2}$cm$^{-3}$ does not produce an extreme inverse
Compton component, which is not observed); Therefore, even if there is
the possibility that a higher density is required, the suitable value
can be derived within the errors only by a broadband modeling of data,
with a consistent variations of all the parameters.

Similarly to the standard model there are univocal correlations
between the estimated spectral parameters and the derived fireball
energy, external density and shock parameters. They are
\be
\tilde{E} \propto \tilde{F_p}^{3/2}\tilde{\nu_m}^{-5/6} \tilde{\nu_c}^{1/4}\tilde{\nu_a}^{-5/6} \delta^{-1/2} T^{0.5}
\label{deltaE}
\ee

\be
\tilde{n}\propto \tilde{F_p}^{-3/2}\tilde{\nu_m}^{25/12} \tilde{\nu_c}^{3/4}\tilde{\nu_a}^{25/6} \delta^{1/2} T^{-7/2}
\label{deltan}
\ee

\be
\tilde{\epsilon_e}\propto \tilde{F_p}^{-1/2}\tilde{\nu_m}^{11/12} \tilde{\nu_c}^{1/4}\tilde{\nu_a}^{5/6} \delta^{1/2} T^{-3/2} 
\label{deltaee}
\ee

\be
\tilde{\epsilon_B} \propto \tilde{F_p}^{1/2}\tilde{\nu_m}^{-5/4} \tilde{\nu_c}^{-5/4}\tilde{\nu_a}^{-5/2}\delta^{-3/2}T^{5/2},
\label{deltaeb}
\ee

\noindent
in slow cooling regime while for a radiative regime 

\be
\tilde{E} \propto \tilde{F_p}^{3/2} \tilde{\nu_c}^{-1/6}\tilde{\nu_a}^{-5/6} \delta^{-1/2}T^{-0.5}
\label{deltaEf}
\ee

\be
\tilde{n}\propto \tilde{F_p}^{-3/2} \tilde{\nu_c}^{17/6}\tilde{\nu_a}^{25/6} \delta^{1/2} T^{7/2}
\label{deltanf}
\ee

\be
\tilde{\epsilon_e}\propto \tilde{F_p}^{-1/2}\tilde{\nu_m}^{1/2} \tilde{\nu_c}^{2/3}\tilde{\nu_a}^{5/6} \delta^{1/2} T^{3/2} 
\label{deltaeef}
\ee

\be
\tilde{\epsilon_B} \propto \tilde{F_p}^{1/2} \tilde{\nu_c}^{-5/4}\tilde{\nu_a}^{-5/2} \delta^{-3/2} T^{-5/2}.
\label{deltaebf}
\ee

\noindent

Some constraints must be added to this set of equations. Because
$\epsilon_e + \epsilon_B + \epsilon_p=1$, where $\epsilon_p$ is the
fractional energy that goes to protons, we most likely expect
$\epsilon_e+ \epsilon_B<0.5$. The kinetic energy in the afterglow
should not be much lower then the energy released in the $\gamma$-ray
phase, otherwise very large efficiencies of radiation, incompatible
with internal shock models, would be required (Lazzati, Ghisellini \&
Celotti 1999).  Finally the apparent source size, the rate of
expansion and the energy in each afterglow should be consistent with
observations in the radio wavebands of interstellar scintillation and
its quenching (Waxman, Kulkarni \& Frail 1998).  In principle these
requirements together with observations of the spectral parameters at
a particular time would set limits on $\delta$ and therefore on the
structure of the magnetic field generated in shocks.  Unfortunately
the current data do not allow a model independent estimation of
spectral parameters.  Actually the spectral shape, and in particular
the break frequencies and the peak flux, are not unequivocally
constrained by data in even the best studied afterglows (e.g. GRB
970508, GRB 000418).  These spectral features are derived by modeling
the sparse measurement collected in various wavebands at different
times: very often different versions of the standard model can fit the
data, consequently finding different spectral parameters (see for
example Berger et al 2000, for GRB 000418).  Moreover these
model dependent quantities (above all $\tilde{\nu_a}$) enter with
higher dependences than $\delta$ in the estimation of the fireball
parameters (Eq.~\ref{deltaE}-\ref{deltaebf}); therefore the break
frequencies and peak flux, derived under the assumption that
$\delta=1$, cannot obviously be used to figure out the value of
$\delta$.  These equations and the physical limits on the fireball
parameters should be used instead to constrain the overall modeling of
data.  The results would actually lead to very interesting
conclusions.

If the modeling succeeded in reproducing the data with $\delta<1$, a
different value for the external density, energy and shock parameters
would possibly be derived. If it indicates a denser environment, this
could be more compatible with the strong observational evidence that
would place GRBs explosions in a star forming region.  However a
\textit{very} short scale magnetic field (say $\delta\lsim 10^{-4}$)
would be probably ruled out, since extremely high densities would be
required in order to get fluxes in the observed ranges and the radio
flux ($\nu<\tilde{\nu_a}$) would probably be too heavily absorbed:
$\tilde{F}_{\nu}\propto \tilde{n}^{-1/2}$ (in adiabatic) and
$\tilde{F}_{\nu}\propto \delta^{-1}\,\tilde{n}^{-5/3}$ (in radiative
regime).  An other implication of the possibility to fit the very same
data with models based on different shock physics assumptions is that
parameter estimations greatly suffer from shock theory uncertainties.
In this work we have only explored the effect of a short-scale
magnetic field, but other reasonable variations of the standard model
(such as time dependent $\epsilon_B,
\epsilon_e $ and  $p$) may affect the parameter estimates as well.
This would call for alternative model independent external density 
and energy estimates in order to constrain,
together with  the afterglow modeling, the shock physics. 

On the other hand if this model can reproduce data only with
$\delta=1$, it could be either evidence in favor of an extremely
stable field or the indication that some other modifications of the
theory must be taken into account.  For example, as already mentioned,
$\delta$ can vary with time: for instance if the Rayleigh-Taylor
instability occurs at the contact discontinuity, the lengthscale would
grow as the fireball evolves and consequently late multibands
observations would be more consistent with $\delta=1$, while early
measurements would require $\delta\ll1$.  Observational tests of this
possibility can be performed using the theory developed in this work;
in fact we can look for a coherent changing in time of $\delta$,
modeling instantaneous spectra at different times, with all the
remaining parameters kept constant. Of, course, with a very good
dataset, this approach can be extended to all the other shock
parameters, including $\epsilon_B, \epsilon_e $ and $p$. Such a
procedure would yield unbiased estimates of shock parameters
variability, and the GRB afterglow would become a laboratory where
shock physic can be tested.  The best case up to now for constraining
$\delta$ and its temporal behavior is probably GRB 000926, but in a
year SWIFT, REM and and possibly the VLA will provide us with more
simultaneous data (also at early times) in X-ray, optical, infrared
and longer wavelengths and therefore more constraining fits will be
performed.  This work actually highlights the importance of observing
afterglows at all wavelengths, from radio to hard x-ray, but
emphasizes as well that current estimations from this data of the
external density surrounding $\gamma$-ray bursts may depend on
uncertain details of shock physics.

\section*{Acknowledgments}
We thank Davide Lazzati for the deep physical insight of his
suggestions and Reem Sari for a very useful critic discussion. We
thank also Andrei Beloborodov for encouraging discussions.  ER thanks
the Isaac Newton and PPARC for financial support.


\end{document}